\documentclass[aip, cha, preprint, author-year]{revtex4-1}

\usepackage{dcolumn}
\usepackage{bm}

\usepackage[utf8]{inputenc}
\usepackage[T1]{fontenc}
\usepackage{mathptmx}
\usepackage{etoolbox}
\usepackage{todonotes}

\usepackage{graphicx} 
\usepackage{wrapfig}
\usepackage{subcaption}
\usepackage{amsmath} 
\usepackage{amsthm}
\usepackage{amssymb} 
\usepackage{comment}
\usepackage{xcolor}
\usepackage{multirow}

\makeatletter
\def\@email#1#2{%
 \endgroup
 \patchcmd{\titleblock@produce}
  {\frontmatter@RRAPformat}
  {\frontmatter@RRAPformat{\produce@RRAP{*#1\href{mailto:#2}{#2}}}\frontmatter@RRAPformat}
  {}{}
}%
\makeatother
\begin{document}


\title{Resilience of the Atlantic Meridional Overturning Circulation}

\author{Valérian Jacques-Dumas}
\affiliation{Institute for Marine and Atmospheric research Utrecht, Department of Physics, 
Utrecht University, Princetonplein 5, 3584 CC Utrecht, Netherlands}
\affiliation{Center for Complex Systems Studies, Department of Physics, 
Utrecht University, Leuvenlaan 4,
3584 CE, Utrecht, Netherlands}

\author{Henk A. Dijkstra}
\affiliation{Institute for Marine and Atmospheric research Utrecht, Department of Physics, 
Utrecht University, Princetonplein 5, 3584 CC Utrecht, Netherlands}
\affiliation{Center for Complex Systems Studies, Department of Physics, 
Utrecht University, Leuvenlaan 4,
3584 CE, Utrecht, Netherlands}

\author{C. Kuehn}
\affiliation{Multiscale and Stochastic Dynamics, Department of Mathematics, Technical University of Munich, Boltzmannstraße 3, 85748 Garching b. München, Germany}

\begin{abstract}
We address the issue of resilience of the Atlantic Meridional Overturning Circulation (AMOC) 
given  the many indications that this dynamical system is in a multi-stable regime. A novel approach to resilience based on rare event techniques is presented which leads to a measure capturing `resistance to change` and `ability to return' aspects in a probabilistic way. The application of this measure to a conceptual model demonstrates its suitability for assessing AMOC resilience but also shows its potential use in many other non-autonomous dynamical systems. This framework is then extended to compute the probability that the AMOC undergoes a transition conditioned on an external forcing. Such conditional probability can be estimated by exploiting the information available when computing the resilience of this system. This allows us to provide a probabilistic view on safe operating spaces by defining a conditional safe operating space as a subset of the parameter space of the (possibly transient) imposed forcing. 
\end{abstract}

\maketitle

\begin{quotation}
The Atlantic Meridional Overturning Circulation (AMOC) is an important component of the climate system due to its role in the meridional heat transport. Theory suggests that it may be in a multi-stable regime and prone to tipping to a collapsed state, which would have dramatic consequences on the climate system. Recent observations and model studies show that this circulation is already weakening. The implications of a collapse of the AMOC call for an assessment of its resilience against the imposed anthropogenic forcing, i.e. the study of its ability to resist change and absorb perturbations. Resilience has attracted increasing interest since its development in the field of ecology but most of its definitions rely on the knowledge of mathematical properties of the studied system. Since these are unknown for very complex and climate models, we develop a new data-driven framework of resilience for the AMOC. So-called "rare event algorithms" have been designed to efficiently sample trajectories containing a rare transition such as an AMOC collapse. We use one of these algorithms to build an ensemble of trajectories in a conceptual AMOC model where the AMOC loses resilience by collapsing. We can then exploit the information contained in this ensemble to characterize the dynamics of the loss of resilience in this model. Furthermore, this framework allows to compute the probability that the AMOC collapses conditioned on the observation of the system at a certain time. From this information, we can deduce a critical boundary, relatable to observable quantities, beyond which the AMOC is assured to collapse. 
\end{quotation}

\section{\label{sec:level1} Introduction}
The Atlantic Meridional Overturning Circulation (AMOC) is a climate subsystem which plays a major role in the meridional heat transport on Earth. It is observed since 2004 by the RAPID array at 26$^\circ$N in the Atlantic and its strength has decreased from 2004 to 2014 with a recovery afterwards \cite[]{Worthington2021}. Other arrays (OSNAP and SAMBA) monitor the AMOC at its northern and southern boundaries, respectively, but the time series are only about 10 years and no trends can be distinguished yet. Observations of the AMOC are strongly motivated by its classification as a potential tipping element in the climate system \cite[]{Lenton2008, McKay2022}, which can undergo a transition in a few decades to a climate disrupting state, also called an AMOC collapse. 

There is now a substantial body of results supporting the idea that the present-day AMOC is in a multi-stable dynamical regime. Since this idea first emerged from a conceptual model \cite[]{Stommel1961}, multi-stable dynamical regimes have been found in a hierarchy of climate models \cite[]{Weijer2019}. The most detailed model where a bistable dynamical regime has recently been found \cite[]{Westen2023} is the Community Earth System Model (version 1). From this study also a physics-based indicator was suggested characterising such a dynamical regime \cite[]{Westen2024}, i.e. the AMOC carried freshwater transport $F_{ovS}$ at its southern boundary (at 34$^\circ$S). Available observations \cite[]{Bryden2011,Garzoli2013} indicate that $F_{ovS} < 0$, which suggests that the AMOC is in this bistable dynamical regime. 

The present-day AMOC is forced by increasing greenhouse gas emissions and by highly varying atmospheric conditions, i.e., heat, freshwater and momentum fluxes. AMOC collapses were likely involved in the last glacial period in so-called Dansgaard-Oeschger events \cite[]{Vettoretti2015}, but no transitions to a collapsed state have been observed over the last 10,000 years. Furthermore, recent studies, based on AMOC reconstructions over the historical period, have indicated that the AMOC may be approaching a tipping point \cite[]{Boers2021, Ditlevsen2023}. Under changing forcing and noise, it is then natural to ask the question on the resilience of the AMOC in particular whether the AMOC has become `less' or `more' resilient over the last decades. 

Resilience is a broad concept which has emerged from the field of ecology \cite[]{Holling1973} but in general it is supposed to describe the capability of a certain system (i) to resist change due to a perturbation and (ii) to be able to return to its original state after such a perturbation. A recent overview of the mathematical definitions of resilience \cite[]{Krakovska2023} shows the many attempts to characterise both its aspects. Examples are local state-based measures, such as those associated with critical slowdown (e.g. a return time, \cite[]{Boulton2022}) and global basin-based measures, for example exit times out of a basin of attraction \cite[]{Arani2021}. Recently, a new system-functioning-based approach has also been proposed  \cite[]{Schoenmakers2021},  which takes into account the fulfilment of the same functioning by different co-existing stable states, as quantified by a characteristic performance range. 

Which resilience concept would be appropriate for the AMOC? Many of the mathematical measures listed in \cite[]{Krakovska2023} require analytical computations which are difficult or impossible to perform in detailed climate models and only few of those measures are adapted to stochastic systems. Since the AMOC is a complex physical system, we do not have to deal  with adaptivity aspects \cite[]{Ramirez2024} such as in ecosystems. From climate models, the multi-stable nature of the AMOC  is expressed by multiple statistical equilibria which there are two `main' equilibria (the AMOC-on and AMOC-off state) intertwined with possible coexisting states \cite[]{Lohmann2024}. This  may be due to what has been called fragmented tipping \cite[]{Bastiaansen2022} and related to convective feedbacks. Hence, the framework of \cite[]{Schoenmakers2021}, where a characteristic performance range of the AMOC could be its meridional heat transport, is an attractive one. 
 
The resilience of the AMOC in a context of climate change has been addressed in detailed climate models \cite[]{Jackson2018} where it was defined as a critical time of exposure to a specific freshwater perturbation for which recovery does not occur anymore. This is a too limited concept as it neither considers the resistance of the AMOC to the imposed perturbation, nor that the recovery may be to a slightly different equilibrium but with a similar heat transport, nor the recovery time. 
In this paper, we develop a novel approach to resilience strongly motivated by the possibility of noise-induced transitions and transitions induced by non-autonomous bifurcations \cite[]{Ashwin2012} in multi-stable stochastic dynamical systems. When working with stochastic systems, the distribution of trajectories has to be taken into account and the resilience measure should then be probabilistic. However, analytical characteristics of this distribution are not computable in general and sampling this distribution with a Monte-Carlo approach would be far too expensive. Instead, we use an algorithm called Adaptive Multilevel Splitting (AMS) \cite[]{Cerou2007}, specifically designed to sample reactive trajectories between two separate domains of a given phase space. Our framework allows measuring resilience along a well-chosen observable of the system, which also provides interpretable insight into the dynamics.

In section \ref{sec:framework}, we  outline this stochastic approach to resilience, present a new resilience measure and show how to compute it in climate models. This approach is applied in section \ref{sec:results} to a conceptual model of the AMOC, for both autonomous and non-autonomous cases. The framework is then extended in section \ref{sec:safe_space} to define the notion of conditional safe operating space. Finally in section \ref{sec:discuss} the results are summarized and possible extensions are discussed. 

\section{Theory}
\label{sec:framework}

\subsection{Characteristics of resilience}
\label{sec:definition}

The AMOC is modelled as a generic stochastic system:
\begin{equation}
\label{eq:system}
    \left\{\begin{array}{l}
    \mathrm{d}\mathbf{x} = [f(\mathbf{x},t)+g(\mathbf{x},t)]\mathrm{d}t + \sigma(\mathbf{x},t)\mathrm{d}W \\
    \mathbf{x}(0) = \mathbf{X}_0
    \end{array}\right.
\end{equation}
where $f:\mathbb{R}^N\times\mathbb{R}\to\mathbb{R}^N$ represents the drift of the system and the stochastic forcing is represented by a process $W$ (of the same dimension as $\mathbf{x}$) and its amplitude is  $\sigma:\mathbb{R}^N\times\mathbb{R}\rightarrow\mathbb{R}^N \times \mathbb{R}^N$. The AMOC is  supposed to be in statistical equilibrium defining a background state $\mathbf{X}_0$. This system is forced by a bounded function $g:\mathbb{R}^N\times\mathbb{R}\rightarrow\mathbb{R}^N$ that may for example represent changes in the radiative forcing. 

We want to capture two important properties  of resilience: conditionality and relativity. Firstly, the resilience of a system is necessarily studied with respect to a certain external forcing and noise amplitude. Their choice thus conditions the measure of resilience. Secondly, after choosing a forcing function and stochastic perturbation, it has to be determined what change in these quantities impacts the resilience of the system. Resilience is in that regard fundamentally relative: it only becomes meaningful in the comparison of different values of resilience. There is no absolute resilience, but rather differences in resilience, across different systems states and different perturbations applied to the same system states. Last but not least, representing the AMOC as a stochastic systems imposes to make all measures of the system probabilistic \cite[]{Arani2021, Ramirez2024}.

The time evolution of a stochastic system is related to the Fokker-Planck equation, which determines the time evolution of the probability density function (PDF) of each variable. Resilience of a stochastic system could be approached through the characteristics of this PDF under a certain forcing. However, if the studied system is of very large dimension (e.g. global climate models), analytical and numerical computations of the PDF become intractable. Moreover, studying the distribution of trajectories by simulating a large ensemble using, for instance, a Monte-Carlo method is also unfeasible due to its high computational cost. The problem becomes more approachable, however, if we only sample trajectories that lose resilience.  
 
As in \cite[]{Schoenmakers2021}, ``losing resilience"  is here considered to be a shift of regions in phase space: loss of resilience is a dynamical process bringing the system from a known region associated with the ``background" state to another ``undesirable" one.  Here, we assume that at least some characteristics of the undesirable region are known. In particular, we suppose that there exists an observable taking different sets of values in the background state of the system and in the undesirable region. For instance, let's focus on the AMOC strength in the North Atlantic. This variable, which is observed at the Atlantic arrays,  would undergo a strong decrease in case of an AMOC collapse and, conversely, observing a strong decrease of this variable would indicate that the AMOC is collapsing. This observable can thus be used as a suitable footprint of the resilience of the AMOC against a collapse. Hence, what is desired is a method that can sample trajectories between specific phase space regions, at a much lower cost than Monte-Carlo simulations.

\subsection{Rare event techniques}
\label{sec:ams}

Adaptive Multilevel Splitting (AMS) is a rare event algorithm designed to sample reactive trajectories between any two separate regions of a given phase space. Let $\mathcal{A}$ and $\mathcal{B}$ be two domains of the phase space, respectively, representing the background region and the undesirable region. We consider a continuous function $\mathcal{O}:\mathbb{R}^N\to\mathbb{R}$. The values of $\mathcal{O}$ corresponding to the background region and the undesirable region are called $\mathcal{O}_\mathcal{A}$ and $\mathcal{O}_\mathcal{B}$, respectively. $\mathcal{O}$ needs to fulfil several conditions to be a suitable observable:
\begin{align}
\nonumber
    & \mathcal{O}(\mathcal{A}) \ne  \mathcal{O}(\mathcal{B}) \\
    & \mathcal{O}^{-1}(\{\mathcal{O}_\mathcal{A}\}) \subset \mathcal{A} \\
\nonumber
    & \mathcal{O}^{-1}(\{\mathcal{O}_\mathcal{B}\}) \subset \mathcal{B} \\
\nonumber
    & \mathcal{O} \mathrm{\ is\ monotonous\ between} \  \mathcal{O}_\mathcal{A} \mathrm{\ and\ } \mathcal{O}_\mathcal{B} 
\end{align}

The first condition imposes that the observable is non-ambiguous: it cannot take the same values in the phase space regions we wish to separate. The second and third conditions impose that the observable is easily readable: the values of $\mathcal{O}$ assigned to the background region and undesirable region cannot refer to other domains of the phase space. They are also important for the non-ambiguity of the observable: if $\mathcal{O}$ takes the value $\mathcal{O}_\mathcal{A}$, then it is clear that the system lies in its background region. Finally, the third condition makes $\mathcal{O}$ interpretable: if $\mathcal{O}$ is increasing (resp. decreasing), the larger (resp. smaller) its values the closer to $\mathcal{B}$. Without loss of generality, we assume from now on that $\mathcal{O}$ is increasing (otherwise consider $-\mathcal{O}$). 

In this framework, if a trajectory is initially in statistical equilibrium in the domain $\mathcal{A}$, losing resilience is equivalent to transitioning towards the domain $\mathcal{B}$. The purpose of a resilience study is thus to characterize a transition from $\mathcal{A}$ to $\mathcal{B}$. 
The AMS algorithm \cite[]{Cerou2007, Rolland2015a, Rolland2015b, Cerou2019, Rolland2022}, presented in Appendix \ref{app:ams}, is designed to sample reactive trajectories among all trajectories obeying (\ref{eq:system}), however unlikely such a transition is. This algorithm was initially designed \cite[]{Cerou2007} to estimate the transition probability between $\mathcal{A}$ and $\mathcal{B}$ but by doing so it also generates an ensemble of $N$ reactive trajectories at a much lower cost than Monte-Carlo simulations. The efficiency of AMS mainly relies on the quality of the score function $\Phi$ \cite[]{Rolland2015a, Lestang_2018, Lucente2022}, which determines at each iteration what trajectories to delete. The optimal score function is called the committor function. Whatever suitable score function is used, the transition probability estimates are always unbiased but the ensemble of trajectories may be biased. It is unbiased if an only if the committor function is used as score function. Methods exist to improve the score function \cite[]{Lucente2022,Jacques-Dumas2023} but their use is out of the scope of the present study and we stick with a simple physics-informed score function. Finally, it is important to note that a suitable score function has properties in common with $\mathcal{O}$ but they are different in the general case. 

\begin{figure}
    \centering
    \includegraphics[width=0.9\textwidth]{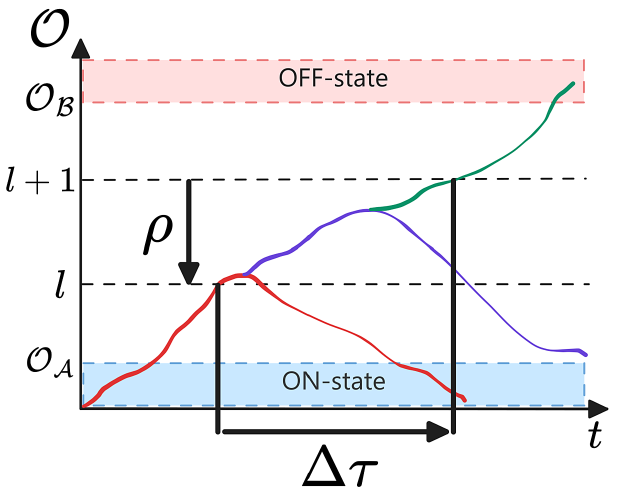}
    \caption{{\it Schematic description of our resilience framework in the AMOC context. The phase space is projected onto the observable $\mathcal{O}$, here the AMOC strength. The AMOC on-state corresponds to all values of $\mathcal{O}\leq\mathcal{O}_\mathcal{A}$, while the off-state of the AMOC is represented by $\mathcal{O}\geq\mathcal{O}_\mathcal{B}$. The dashed lines represent two consecutive isolevels of $\mathcal{O}$. Three trajectories are shown, each branched from the former, as would happen in three consecutive AMS iterations. The number of AMS iterations needed to have them all cross the gap between isolevels $l$ and $l+1$ directly relates to $\rho(\mathcal{O})$. The average time all trajectories spend between these two consecutive levels gives $\Delta\tau(\mathcal{O})$.}}
    \label{fig:sketch}
\end{figure}

The use of AMS within our framework is summarized in Fig. \ref{fig:sketch}. It represents three branched trajectories projected onto the observable $\mathcal{O}$, transitioning from the on-state to the off-state. The dashed lines represent the isolines of the observable. With AMS, all trajectories are simulated from $\mathcal{A}$ until they reach $\mathcal{B}$ or fall back on $\mathcal{A}$. As a result, AMS returns a transition probability (given by eq. \ref{eq:proba}) conditioned on the fact that a given reactive trajectory reaches $\mathcal{B}$ before returning to $\mathcal{A}$. Moreover, this probability only depends on the number of iterations performed by AMS. Since a trajectory going from $\mathcal{A}$ to $\mathcal{B}$ must go through all intermediate values of $\mathcal{O}$ between $\mathcal{O}_\mathcal{A}$ and $\mathcal{O}_\mathcal{B}$ (dashed lines in Fig. \ref{fig:sketch}), the equation (\ref{eq:proba}) also gives the probability to reach any intermediate level $l$ of $\mathcal{O}$ before returning to $\mathcal{A}$. This probability is called $P(l)$ (see Appendix \ref{app:ams} for details on its computation). Bayes' theorem then gives the probability to reach the level $l+1$ before returning to $\mathcal{A}$, given that the level $l$ has already been reached before returning to $\mathcal{A}$ as:
\begin{equation}
    P(l+1\ |\ l) = \frac{P(l\ |\ l+1)P(l+1)}{P(l)} = \frac{P(l+1)}{P(l)}
\end{equation}
because if $\mathcal{O}$ is monotonous, all trajectories that have reached the level $l+1$ have necessarily reached the level $l$ before, hence $P(l\ |\ l+1)=1$. In Fig. \ref{fig:sketch}, we indicate $\rho(\mathcal{O})=1-P(l+1\ |\ l)$, which is the probability that a trajectory will return to $\mathcal{A}$ after reaching $l$ and before reaching $l+1$, thus describing the ability of the system to return to $\mathcal{A}$ at any point of the transition.

Another footprint can be derived from AMS for the second important characteristic of resilience: resistance to the applied forcing. This property is related to the gradient of the mean first-passage times (MFPT) across successive levels of the observable. The MFPT across a given level $l$ can be measured directly in the generated ensemble (see Appendix \ref{app:ams} for details on its computation). This quantity is represented in Fig. \ref{fig:sketch} by $\Delta\tau$ and expressed as a time per observable unit (or simply a time if $\mathcal{O}$ has been normalized and is dimensionless). 

The two footprints of resilience, i.e.:
\begin{equation}
\label{eq:return}
    \rho(\mathcal{O}) = 1 - \frac{P(l+1)}{P(l)} ~ ; ~ \Delta\tau(\mathcal{O}) =\frac{\partial\tau}{\Delta\mathcal{O}} 
\end{equation}
both somehow measure the difficulty of crossing the gap between consecutive levels of $\mathcal{O}$:
\begin{itemize}
    \item $\rho$ is related to the probability that a trajectory returns to its background state while the observable $\mathcal{O}$ lies between $\mathcal{O}_l$ and $\mathcal{O}_{l+1}$
    \item $\Delta\tau$ describes the time that a trajectory would take in average to cross the gap between these two levels.
\end{itemize}
However, if the system has to climb a potential wall, $\rho(l)$ will surge for all levels $l$ belonging to the wall because the probability of going over it is very small ; but a trajectory that can pass it will likely pass it quite fast, hence $\Delta\tau(l)$ will be small. Both footprints thus provide complementary information about the behavior of the system. 

One of the main advantages of this method, is that these two important footprints of resilience, $\rho$ and $\Delta\tau$, can be measured at no additional cost than that of running AMS, since they are sampled from the generated trajectories at every iteration of this algorithm. Moreover, these footprints respect the conditionality of resilience, since their values are conditioned on the chosen forcing function and noise amplitude.  

\subsection{Resilience measure}
\label{sec:measure}

The two resilience footprints in (\ref{eq:return}) can be transformed into a single indicator that makes it easier to measure and characterise the resilience of the AMOC. A curve in the $(\Delta\tau,\rho)$ plane (Fig.~\ref{fig:sketch}) displays the evolution of the resilience of the given system when transitioning between its background region (one end of the curve) to an undesirable region (other end of the curve). Along this curve, any decrease in $\rho$ indicates a drop in the ability of the system to return to its background region, thus the crossing of some critical threshold. Conversely, any increase in $\rho$ relates to an increase in the resilience of the system, since it may avoid the transition. With the same reasoning, any decrease in the value of $\Delta\tau$ means that the system is accelerating towards the undesirable regime (less time spent between every consecutive levels) and thus losing resilience. Conversely, an increase in $\Delta\tau$ means that the system is slowing down, which increases opportunities to act on it to prevent the transition, thus increasing its resilience. 

Resilience can then naturally be thought as a distance in the $(\Delta\tau,\rho)$ diagram: the larger the distance from the origin, the larger the resilience of the system (given a certain forcing and noise amplitude). A straightforward choice of distance to measure resilience in this plane would be the Euclidean distance. $\rho$ can only take values between $0$ and $1$ (it is a probability) and all values of $\Delta\tau$ are normalized to their maximum value (in the ensemble generated by one AMS experiment). This means that for a given distance $r$ to the origin in this diagram, the same resilience value is given to two limiting cases: $(\Delta\tau=0,\rho=r)$ and $(\Delta\tau=r,\rho=0)$. However, these situations are not symmetrical: in the first case, the system is facing a potential wall, which can be overcome very fast but this event is extremely unlikely to happen (if $r$ is large) and in the second case, the system is drifting very slowly towards the undesirable region but has almost no chance to return to its background state. In this situation, the ability to return is a more meaningful indicator of resilience, because as it tends to zero, it becomes almost certain that the system is bound to completely lose resilience. To emphasize the asymmetry between $\rho$ and $\Delta\tau$, we modify the Euclidean distance in the $(\Delta\tau,\rho)$ plane by multiplying it by $\rho$.  

Finally, we consider the resilience of the system to be determined by the whole history of the transition between regions, leading to the integral measure:
\begin{equation}
\label{eq:resilience}
    {\cal R} = \int_{\mathcal{O}_\mathcal{A}}^{\mathcal{O}_\mathcal{B}} \rho(\mathcal{O})\sqrt{\rho(\mathcal{O})^2+\Delta\tau(\mathcal{O})^2}\mathrm{d}\mathcal{O}
\end{equation}
This measure of resilience is bounded below by $0$, corresponding to an instantaneous transition from the background region to the undesirable region and thus an absence of resilience. It is theoretically bounded above by $(\mathcal{O}_\mathcal{B}-\mathcal{O}_\mathcal{A})\sqrt{2}$ in the case where the system transitions at a constant speed while being assured to return to its background region at all times. This (ideal) case corresponds to the largest possible resilience.  

This resilience indicator is easily interpretable: a larger value of ${\cal R}$ corresponds to a larger resilience of the system. It takes into account the two important aspects of resilience: ability to return and resistance to change. Moreover, it respects two characteristics of resilience: ${\cal R}$ is conditioned on the forcing and the noise imposed on the system. It is also conditioned on the choice of background and undesirable regions, although these may be straightforward in the case of the AMOC. Finally, this indicator is relative: a value of ${\cal R}$ does not bear any meaning per se, but always has to be compared to another value of ${\cal R}$ corresponding to a different forcing (or noise). 
The indicator ${\cal R}$ partly relates to the mean-first passage time, which is a well-known resilience indicator, while acknowledging that it is not the only important variable. Our framework to determine $\mathcal{R}$ relies on prognostic measures with a predictive power and, furthermore, the analysis of the evolution of $\rho$ and $\Delta\tau$ are still possible during a transition and can be used to highlight the behaviour of the system. 

\section{Results}
\label{sec:results}

\subsection{The conceptual AMOC model}

\begin{figure*}[t]
    \centering
    \includegraphics[width=\linewidth]{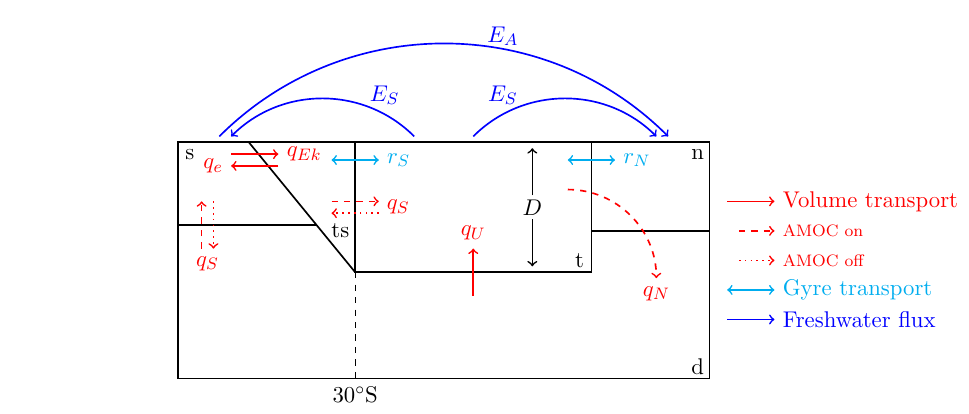}
    \caption{{\it The AMOC model described by \cite[]{Castellana2019}. Full lines represent fluxes that are always present in the model. Dashed lines represent the circulation in the present-day stable state. Dotted lines represent the weaker, reversed circulation in the collapsed stable state. The noise and the freshwater forcing prescribe the flux $E_A$.}}
\label{fig:model}
\end{figure*}

The model of the AMOC used here was first described by \cite[]{Cimatoribus2014} and later extended by \cite[]{Castellana2019}, whose setup we reproduce here. This model consists of five boxes (Fig. \ref{fig:model}), each representing a region of the Atlantic Ocean: the northern Atlantic (labelled n), the southern Atlantic (labelled $s$), the deep ocean (labelled d), the tropical Atlantic (labelled t) and a southern tropical region (labelled ts). The only variables are the salinity of each box and the depth of the pycnocline $D$. There are three main volume transport terms between the boxes that drive the overturning circulation: $q_S$, corresponding to the transport in the southern Atlantic, $q_N$, describing the downwelling in the northern Atlantic, and $q_U$, which describes the upwelling across the pycnocline. $q_N$ corresponds to the AMOC strength and we also use it as observable. 

The forcing is represented by a freshwater flux from the southern to the northern Atlantic, decomposed into a fixed symmetric flux ($E_s$) and a variable asymmetric flux $E_A(t) = \overline{E_A}(1+f_\sigma\zeta(t))$. $\zeta$ represents a zero-mean, unit-variance white noise process and $\overline{E_A}$ and $f_\sigma$ are two control parameters: the former controls the amplitude of the forcing and the second the amplitude of the noise.

When $\overline{E_A}\in[0.06,0.35]\ $Sv ($1\ \mathrm{Sv}=10^6\ \mathrm{m}^3\mathrm{s}^{-1}$), the AMOC in this model is in a bistable regime. One of the stable states corresponds to the present-day circulation, with $q_N>q_S>q_U>0$. The other stable state represents a collapsed circulation in the northern Atlantic and a weakened reversed circulation in the south: $q_N=0$ and $q_S<0$. However, there exists an region in the phase space corresponding to a temporary shut-down of the circulation, with $q_N=0$ and $q_S>0$. We are mostly interested in transitions from the current state of the AMOC to this second type of collapse because its timescale is of the order of $100$ years, as opposed to $1000$ years for the transitions between both stable states. 

\subsection{Resilience of the AMOC}

\subsubsection{Autonomous setting}
\label{sec:auton}

We follow the procedure described in Sect. \ref{sec:ams} and Sect. \ref{sec:measure} and apply the AMS algorithm to the AMOC model. We generate an ensemble of $N=1000$ trajectories losing resilience, that is transitioning from the stable state corresponding to the present-day AMOC (background state) to the domain in phase space where $q_N=0$ (temporary shutdown which is the undesirable state). At each iteration of AMS, $n_c=10$ trajectories are deleted. The score function $\Phi$ is a simple function of the AMOC strength $q_N$, as in \cite[]{Castellana2019}:
\begin{equation}
\label{eq:score}
    \Phi(\mathbf{x}) = 1 - \frac{q_N(\mathbf{x})}{q_0}
\end{equation}
where $q_0$ represents the AMOC strength of the present-day stable state. This same function is taken as observable in AMS. It is equivalent to using the AMOC strength as observable, but is conveniently increasing and bounded between $0$ and $1$. The background region is chosen as: $\mathcal{A} = \{\mathbf{x}\ |\ q_N(\mathbf{x})\geq q_0\}$. The undesirable region is: $\mathcal{B} = \{\mathbf{x}\ |\ q_N(\mathbf{x})=0\}$. These domains are well-separated in phase space, as shown by $\Phi$, which is equal to $0$ in $\mathcal{A}$ and to $1$ in $\mathcal{B}$. 
We run AMS for $36$ values of $\overline{E_A}$ spanning the interval $[0.03,0.38]\ $Sv, so as to cover a slightly larger interval than the bistability regime. We consider $4$ noise amplitudes, from $\overline{E_A}f_\sigma=0.02\ $Sv to $\overline{E_A}f_\sigma=0.05\ $Sv, where the values of $f_\sigma$ vary with the different values of $\overline{E_A}$ to keep $\overline{E_A}f_\sigma$ constant. AMS is run $30$ times for each couple $(\overline{E_A},f_\sigma)$. 

\begin{figure}
    \centering
    \includegraphics[width=\textwidth]{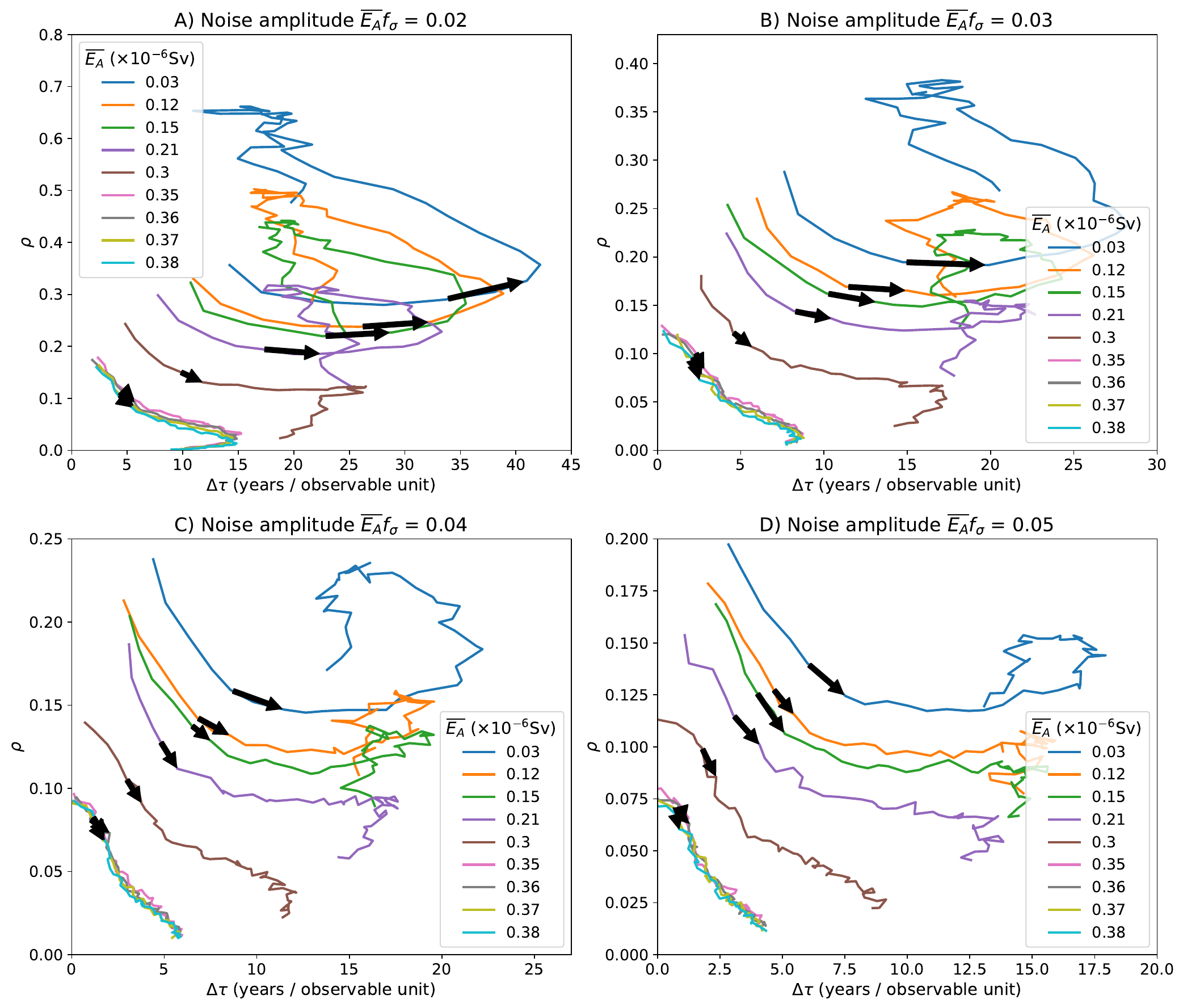}
    \caption{{\it $(\Delta\tau,\rho)$ diagrams representing the evolution of the resilience footprints while the AMOC transitions from $\mathcal{A}$ to $\mathcal{B}$, for different values of noise amplitude. Each panel A-D represent the results associated to a different noise amplitude. Each curve represents the mean values of $\rho$ and $\Delta\tau$ across the $30$ runs of AMS. The black arrows indicate the direction of time. Overall, in all cases, a smaller forcing ($\overline{E_A}$) results in larger values of $\Delta\tau$ and $\rho$. For $\overline{E_A}\geq0.35\times10^{-6}\ $Sv, all curves are identical, which shows the change of dynamics induced by the saddle-node bifurcation.}}
    \label{fig:diagram_auton}
\end{figure}

The $(\Delta\tau,\rho)$ diagrams for each of the noise amplitudes are shown in Fig. \ref{fig:diagram_auton}. Each diagram allows for an analysis of the qualitative behaviour of the system while it is losing resilience. Note in all panels of Fig. \ref{fig:diagram_auton} that all curves exhibit ``stochastic-looking" features, especially for the smallest values of $\overline{E_A}$. These are direct consequences of the AMS process: during each AMS run, $\Delta\tau$ is averaged over a finite-sized collection of stochastic trajectories. The quantity $\rho$ also results from the behaviour of this collection of stochastic trajectories and the numerical spread of both $\Delta\tau$ and $\rho$ strongly depends on the quality of the score function. Both $\rho$ and $\Delta\tau$ are random variables, one sample of which is obtained with each AMS run. Their values presented in Fig. \ref{fig:diagram_auton} are averaged over $30$ runs of AMS. We have already studied this model and showed \cite[]{Jacques-Dumas2024} that the score function $\Phi$ (eq. \ref{eq:score}) is close enough to the committor function so that the confidence interval around the probabilities (and hence $\rho$) is quite narrow and we can be confident that the sampled trajectories are not biased. Overall, the errorbars on both quantities increase as the noise amplitude $\overline{E_A}f_\sigma$ decreases and as the observable tends to $1$ (i.e. when trajectories get closer to the collapse threshold). This is consistent with AMS being less precise as the transition probabilities become very small, which is the case when the noise is small. As a result, the overall shape of the curves in Fig. \ref{fig:diagram_auton} is reliable but the small-scale structures are difficult to interpret. 

As the forcing amplitude $\overline{E_A}$ increases, the probability of return to the background state and the resistance of the system decrease overall, which is to be expected. Moreover, all curves for $\overline{E_A}\geq0.35$ are identical. This value of the freshwater forcing corresponds to the saddle-node bifurcation where the AMOC on-state disappears \cite[]{Castellana2019}. As a result, the system collapses very fast, which is captured by the small values of $\Delta\tau$ and the drop in $\rho$. First consider Fig. \ref{fig:diagram_auton}A. For $\overline{E_A}=0.3\times10^{-6}\ $Sv, $\Delta\tau$ first increases much faster than $\rho$ decreases. Then, both $\rho$ and $\Delta\tau$ decrease together. We can interpret this as a slow-down of the system as it is moving towards a critical domain in phase space beyond which return to the background state is much less likely. Once this area has been crossed, the system accelerates again to finally collapse soon after. An interesting change of dynamics occurs for $\overline{E_A}\leq0.21\times10^{-6}$Sv. Once again, the system starts by slowing down while keeping $\rho$ constant. Then, $\Delta\tau$ and $\rho$ suddenly increase together, meaning that the system accelerates again while having an increased chance of returning. Finally, the increase in $\rho$ levels off and $\Delta\tau$ increases again while $\rho$ drops. As $\overline{E_A}$ decreases, the stability landscape evolves and the aforementioned critical area may have become a ``potential wall": the probability of return increases while the trajectory is climbing the wall (the difference in probability between each successive level on the wall becomes larger) and drops when the wall has been passed. Furthermore, decreasing $\overline{E_A}$ increases the relative importance of the noise, which may result in a faster crossing of the "wall". If the rest of the stability landscape is overall flat, the system may then slow down again. As $\overline{E_A}$ is reduced again, the effect of this ``potential wall" becomes more important, to the point where the system barely undergoes a drop in $\rho$ before collapsing for $\overline{E_A}=0.03\times10^{-6}$Sv. The dynamics observed for a larger noise amplitude in Fig. \ref{fig:diagram_auton}B are qualitatively the same. However they change in Fig. \ref{fig:diagram_auton}C, where, even at the beginning of the trajectories, $\rho$ drops as $\Delta\tau$ increases. For the smallest values of $\overline{E_A}$, a critical area can still be found where $\rho$ increases, but this increase does not come with a significant decrease in $\Delta\tau$ which characterised (for smaller noise cases) a ``potential wall". Now, the noise becomes large enough to flatten the stability landscape. Increasing the amplitude of the noise in Fig. \ref{fig:diagram_auton}D only exacerbates this effect.

\begin{figure}
    \centering
    \includegraphics[width=0.7\linewidth]{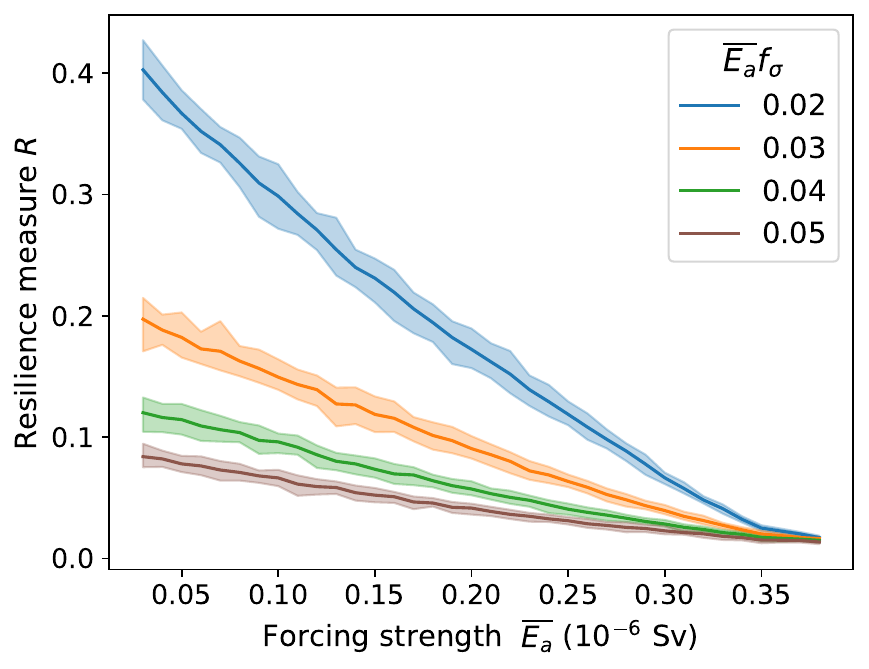}
    \caption{{\it Resilience measures ${\cal R}$ of the AMOC for different values of $(\overline{E_A},f_\sigma)$. The solid line corresponds to the average computed over $30$ AMS runs and the shaded areas show the corresponding $90\%$ confidence intervals. The resilience indicator depends linearly on $\overline{E_A}$ but increases exponentially as $\overline{E_A}f_\sigma$ decreases. The saddle-node is visible for $\overline{E_A}=0.35\times10^{-6}$Sv as a change of slope in all four curves.}}
    \label{fig:measure_auton}
\end{figure}

The resilience measure $\mathcal{R}$ of the system under the chosen constraints for the parameters $(\overline{E_A},f_\sigma)$ is presented in Fig. \ref{fig:measure_auton}. The solid lines correspond to the mean $\mathcal{R}$ computed over the $30$ AMS runs (for each couple $\overline{E_A},f_\sigma$) and the coloured areas show the corresponding $90\%$ confidence intervals. These measures are in accordance with the finding in Fig. \ref{fig:diagram_auton} that $\rho$ and $\Delta\tau$ overall increase as $\overline{E_A}$ decreases, leading to increased values of ${\cal R}$, interpreted as larger resilience. Furthermore, $\mathcal{R}$ grows linearly as a function of the prescribed freshwater forcing. Its slope can be quantified by performing a linear regression on the values of resilience against $\overline{E_A}$. The R-squared values and slopes for $\overline{E_A}\leq0.35\times10^{-6}\ $Sv are presented in Tab. \ref{tab:slopes}. The saddle-node bifurcation at $\overline{E_A}=0.35\times10^{-6}\ $Sv is visible in Fig.~\ref{fig:measure_auton} as a breaking in the linearity of $\mathcal{R}$: it still has a linear behaviour for $\overline{E_A}\geq0.35\times10^{-6}\ $Sv, but with a smaller slope.  

\begin{table}
    \centering
    \begin{tabular}{c|c|c}
    $\overline{E_A}f_\sigma$ & $R$-squared & Slope \\
    \hline
    0.02 & 0.993 & -1.17 \\
    0.03 & 0.995 & -0.57 \\
    0.04 & 0.990 & -0.33 \\
    0.05 & 0.987 & -0.22 \\
    \end{tabular}
    \caption{{\it $R$-squared values and slopes of all four curves displayed in Fig. \ref{fig:measure_auton}. The $R$-squared show that the resilience indicator behaves indeed linearly with $\overline{E_A}$ for $\overline{E_A}\leq0.35\times10^{-6}\ $Sv. The slopes are useful to quantify the loss of resilience of the system with an increase of the freshwater forcing.}}
    \label{tab:slopes}
\end{table}

For $\overline{E_A}\leq0.35\times10^{-6}\ $Sv, the value of the R-squared coefficients show that the linear approximation is robust, all the more so as the noise amplitude decreases. Tab. \ref{tab:slopes} provides a quantification of the loss a resilience of the AMOC depending on the freshwater forcing. However, the large difference between the slope corresponding to $\overline{E_A}f_\sigma=0.02$ and that corresponding to $\overline{E_A}f_\sigma=0.03$ shows that this model is most vulnerable to an increase of its natural variability. Increasing the noise amplitude from $0.02$ to $0.03$ is the most efficient way of decreasing the resilience of the system against freshwater forcing. \\

\subsubsection{Non-autonomous setting}
\label{sec:non_auton}

The same procedure can be used when the forcing applied to the system is non-autonomous. This case is especially of interest when considering climate change, due to the time-dependence of the anthropogenic forcing. Here, we consider a simple ramp-up forcing, where $\overline{E_A}$ increases from $0.15\ $Sv to a certain target value $\overline{E_A}_f$. We study two different target values: $\overline{E_A}_f=0.25\ $Sv and $\overline{E_A}_f=0.35\ $Sv. In the first case, the ramp-up times range from $10$ to $100$ years and in the second case from $10$ to $200$ years. 
Whatever the value of $\overline{E_A}_f$, a constant noise amplitude $\overline{E_A}f_\sigma=0.02\ $Sv is used (so effectively $f_\sigma$ is decreased during the ramp-up of $\overline{E_A}$). When applying AMS in such non-autonomous setting, extra caution is needed for the cloning of trajectories: the branching time matters because it directly impacts the forcing of the newly-computed trajectory. 

\begin{figure}
    \centering
    \includegraphics[width=\textwidth]{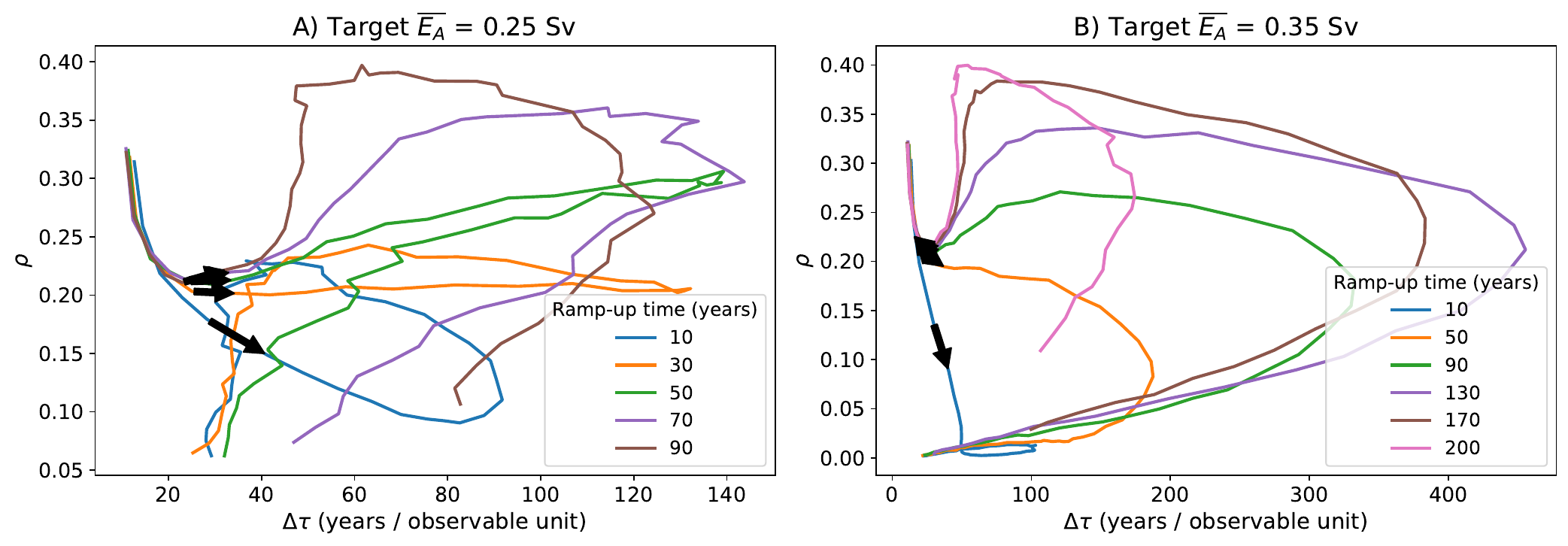}
    \caption{{\it $(\Delta\tau,\rho)$ diagrams representing the evolution of the resilience footprints while the AMOC model is transitioning from $\mathcal{A}$ to $\mathcal{B}$ under an increasing forcing, for several ramp-up times. Each panel A-B represent the results associated to a different target value $\overline{E_A}_f$, while the starting value of $\overline{E_A}$ is fixed to $0.15\ $Sv. All AMS experiments were run for a constant noise $\overline{E_A}f_\sigma=0.02\ $Sv. Each curve represents the mean values of $\rho$ and $\Delta\tau$ across the $30$ runs of AMS. The black arrows indicate the direction of time.}}
    \label{fig:diagram_non_auton}
\end{figure}

The resilience analysis provided by AMS in this non-autonomous setting is presented in Fig. \ref{fig:diagram_non_auton}. The curves in these diagrams can be compared to the curve in Fig. \ref{fig:diagram_auton}A corresponding to a constant $\overline{E_A}=0.15\ $Sv with the same noise amplitude of $0.02\ $Sv. In both panels of Fig. \ref{fig:diagram_non_auton}, as in the autonomous case, all curves show the same initial decrease in $\rho$ while $\Delta\tau$ remains constant because $\overline{E_A}$ remains close enough to $0.15\ $Sv. The system is initially escaping at constant speed from the immediate neighbourhood of the AMOC on-state. After this initial decrease of $\rho$, the non-autonomous effects become visible. 

As the ramp-up time of the forcing increases, the curves in both panels of Fig. \ref{fig:diagram_non_auton} show a similar behavior. For the fastest ramp-up, $\rho$ keeps decreasing, so the system quickly loses resilience. As the forcing rate decreases, this decrease of $\rho$ slows down and even reverts: for a large enough ramp-up time (about $50$ years in panel A and $90$ years in panel B), $\rho$ increases again after its initial drop. As the ramp-up time increases, the amount of AMS iterations needed to make trajectories pass consecutive levels increases, which means that, as expected, the slower the forcing rate the more difficult for the system to transition. However, in all curves, the system ends up passing a threshold beyond which $\rho$ drops until the AMOC collapses, showing that the AMOC reached a critical point beyond which returning to its initial state is increasingly difficult. Ramping-up $\overline{E_A}$ for longer also induces a slowing down of the collapse. All curves of Fig. \ref{fig:diagram_non_auton} reach a maximum value of $\Delta\tau$ of at least $90$ years/observable unit, which is much larger than the largest value of $\Delta\tau$ reached by any curve in the autonomous case (see Fig. \ref{fig:diagram_auton}, where the scale of $\Delta\tau$ reaches at most $45$ years/observable unit). The ramping up of the forcing creates a slowdown of the system, which is more pronounced when the ramp-up takes longer and when the target value of $\overline{E_A}$ is larger (see the scales of $\Delta\tau$ in panels A and B: in panel B, the maximum value reached of $\Delta\tau$ is about $3$ times larger than in panel A). Trajectories thus spend a longer time between two consecutive levels as the forcing rate decreases. But once again, the system always reaches a threshold beyond which $\Delta\tau$ drops, so the system accelerates until the collapse. 

\begin{figure}
    \centering
    \includegraphics[width=0.7\textwidth]{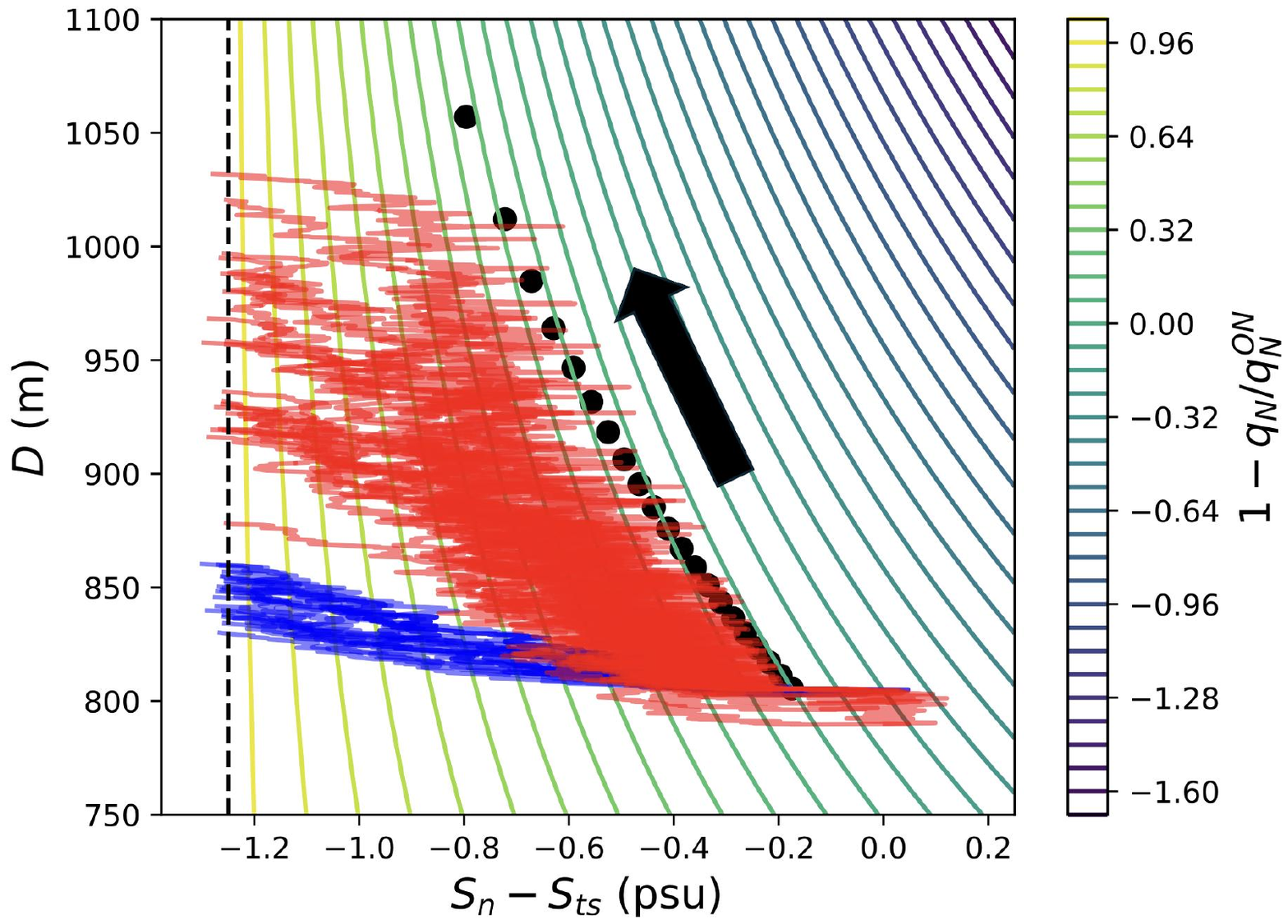}
    \caption{{\it Salinity difference $S_n-S_{ts}$ between boxes n and ts against the pycnocline depth $D$. The blue and red trajectories correspond, respectively, to the autonomous and non-autonomous case. In the autonomous case, the forcing $\overline{E_A}=0.15\ $Sv is constant. In the non-autonomouse case, the forcing starts at this value and is ramped-up for $100$ years until $0.35\ $Sv. The dashed vertical line represents the threshold of the AMOC collapse and the dots correspond to the AMOC on-state for $\overline{E_A}\in[0.15,0.35]\ $Sv. The arrow shows the change in its position as $\overline{E_A}$ increases. The colored contours represent the isolevels of the normalized strength of the AMOC, used here as score function for AMS and as observable. It is equal to $0$ at the on-state for $\overline{E_A}=0.15\ $Sv and to $1$ at the collapse.}}
    \label{fig:phase_space}
\end{figure}

Fig. \ref{fig:phase_space} helps understanding this slowdown. It represents the salinity difference between boxes n and ts plotted against the pycnocline depth ; the fixed points in this model only depend on those quantities. All trajectories start from the AMOC on-state and stop when reaching the dashed line, beyond which the downwelling in the North Atlantic vanishes. Blue trajectories represent the autonomous case (green curve in Fig. \ref{fig:diagram_auton}A) while red trajectories represent the non-autonomous case of Fig. \ref{fig:diagram_non_auton}B with a ramp-up time of $100$ years. As $\overline{E_A}$ increases, the AMOC is pushed towards collapsing but also pulled back to the stable on-state. Since this state moves roughly along the isolines of the observable, it takes more time to cross them, hence the observed slowdown. Finally, as the ramp-up time increases, it also reaches a threshold, beyond which it is too slow and the system accelerates again, which translates to a decrease in the values of $\Delta\tau$ reached during the transition. This can be seen in Fig. \ref{fig:diagram_non_auton}A for a ramp-up time of $90$ years and in Fig. \ref{fig:diagram_non_auton}B for a ramp-up time of $170$ and $200$ years. For an even smaller rate of forcing, the behaviour of the system in the $(\rho,\Delta\tau)$ diagram looks like its behaviour in the autonomous case.

\begin{figure}
    \centering
    \includegraphics[width=0.7\linewidth]{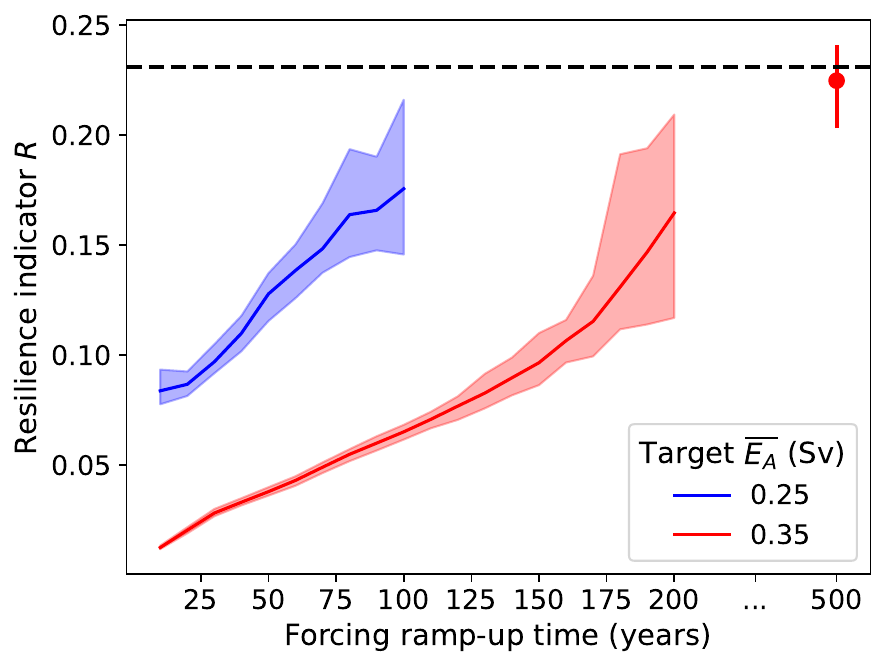}
    \caption{{\it Resilience measures ${\cal R}$ of the AMOC for two cases of the ramp-up, that only differ by the target value of $\overline{E_A}$. Each solid line corresponds to the average of $\mathcal{R}$ computed over $30$ AMS runs and the shaded areas show the corresponding $90\%$ confidence intervals. The dashed line represents the value of $\mathcal{R}$ in the autonomous setting for $\overline{E_A}=0.15\ $Sv and a noise amplitude of $\overline{E_A}f_\sigma=0.02\ $Sv. The red dot represents the resilience indicator in the non-autonomous for a ramp-up time of $500$ years and is plotted with its $90\%$ confidence interval.}}
    \label{fig:measure_non_auton}
\end{figure}

The resilience indicator $\mathcal{R}$ in this non-autonomous setting is presented in Fig. \ref{fig:measure_non_auton}. As expected, it increases with the ramp-up time, while staying below its value in the autonomous case (dashed line). This is expected because the larger forcing applied in the non-autonomous setting triggers a decrease (compared to the autonomous case) in $\rho$, taken as driving variable of the resilience indicator. For the same reason, we find in Fig. \ref{fig:measure_non_auton} that the larger $\overline{E_A}_f$, the smaller $\mathcal{R}$. Note the very large $90\%$ confidence interval around the resilience indicator as the ramp-up time increases, especially with a target value $\overline{E_A}_f=0.35\ $Sv. In the autonomous case, the width of this confidence interval (see Fig. \ref{fig:measure_auton}) was about $15\%$ of the mean resilience indicator. Here, for a target $\overline{E_A}=0.25\ $Sv, the width of the confidence interval is at most $40\%$ of the mean resilience indicator and even $60\%$ and for a target $\overline{E_A}_f=0.35\ $Sv. This is due to the amplification of the variance of the MFPTs on the trajectories sampled with AMS, as can be seen in Fig. \ref{fig:phase_space}. The slower the ramp-up, the more trajectories are prone to collapse as in the autonomous case (blue ensemble in Fig. \ref{fig:phase_space}). However, other trajectories follow the movement of the on-state in phase space and take on average longer to collapse. This increased variance in the length of trajectories is exacerbated by the cloning of random trajectories during the AMS process. Note that this large variance on $\mathcal{R}$ is a feature of the studied system due to the drift of its on-state with $\overline{E_A}$, rather than an consequence of the resilience framework. 
If the ramp-up time increasing again, it is expected that the variance on the MFPTs will not keep increasing and most trajectories will behave as in the autonomous case. This is shown by the red dot in fig. \ref{fig:measure_non_auton}, computed with a ramp-up time of $500$ years. Its confidence interval overlaps with the value of $\mathcal{R}$ in the autonomous case and the relative width of its confidence interval is similar as in the autonomous case.

\section{Conditional safe operating spaces}
\label{sec:safe_space}

Safe operating spaces \cite[]{Rockstrom2009,Steffen2018} have gained increasing attention in the fields of resilience and sustainability. They were originally defined as the regime of different climate subsystems under "planetary boundaries" \cite[]{Rockstrom2009}, which are thresholds (surrounded by a zone of uncertainty) beyond which a given climate subsystem would undergo a tipping or significant change at least at the continental level. In this view, a safe operating space is the region either in phase space or parameter space where a climate system must remain to prevent a disruption of human presence or life processes. Here, we extend our resilience framework to define a probabilistic view of safe operating spaces. 

Let $E$ be the event of interest with a certain probability of occurrence $P(E)$. Here, $E$ represents a direct collapse of the AMOC (transition between the on-state $\mathcal{A}$ and the off-state $\mathcal{B}$ without returning to the on-state) as sampled by AMS. 
Let $C(\{a_i\}_{i\in[1,k]})$ be a class of events depending on parameters $\{a_i\}_{i\in[1,k]}$. In principle, $C$ could represent anything, such as another climate subsystem interacting with the system of interest. Let $C(l,t)$ designate the crossing of the level $l$ of $\mathcal{O}$ before a time $t$ by a trajectory starting from the AMOC on-state (and before returning to this on-state). The goal is to study how $C(\{a_i\}_{i\in[1,k]})$ impacts the probability of occurrence of $E$, as a function of $\{a_i\}_{i\in[1,k]}$. Measuring the realization of any event in the class $C$ then provides information on the state of the AMOC and potentially an early warning signal that it is endangered, if $P(E\ |\ C(\{a_i\}_{i\in[1,k]}))>P(E)$. The question then is: if the current strength of the AMOC is measured and found to have passed a certain threshold, how does that inform us about the probability of a direct collapse of the circulation?

A safe operating space for $E$ would correspond to the set of physical states or system parameters ensuring that the probability that the AMOC collapses is not (dramatically) increased under a specific forcing or noise amplitude. This safe operating space is not tractable in practice due to its too large dimensionality: the number of variables and parameters to consider is huge in principle. Considering the safe operating space of $E$ conditioned on $C$ focuses the problem on the impact that the realization of $C$ would have on the realization of $E$. In our example, this conditional safe operating space only has two dimensions, which are the two parameters $l$ and $t$ of $C$ and is described by all values of $P(E\ |\ C(l,t))$.
As in Sect. \ref{sec:ams}, this conditional probability is computed using the Bayes' theorem:
\begin{equation}
    P(E\ |\ C(l,t)) = \frac{P(C(l,t)\ |\ E)P(E)}{P(C(l,t))}
\end{equation}

First, $P(E)$ is assumed to be known because it has been computed with AMS. Similarly, $P(C(l,t))$ can be computed with AMS. $P(C(l,t)\ |\ E)$ can be estimated by exploiting information already available. Indeed, computing $P(E)$ with AMS implies that we obtain an ensemble of trajectories all realizing $E$. We can then count in these trajectories how many times the events $C(l,t)$ are realized to estimate $P(C(l,t)\ |\ E)$. This estimate is of course quite rough, strongly depends on the number of AMS runs of $P(E)$ and on the size of the ensemble used during these runs. The main advantage is that this estimate is automatically available at no additional cost as long as $P(E)$ is computed with AMS.

In applying this to the AMOC model, we consider the forcing scenario to be fixed (both for the computation of $P(E)$ and $P(C(l,t)$): $\overline{E_A}$ is ramped up for $100$ years, from $0.15\ $Sv to $0.35\ $Sv. The noise amplitude is fixed to $\overline{E_A}f_\sigma=0.02\ $Sv. The parameter $l\in[0,1]$ spans the whole range of the observable $\mathcal{O}=\Phi(\mathbf{x})$ (see Sect. \ref{sec:auton}), where $0$ represents the AMOC on-state and $1$ its collapsed state. Twenty equally spaced samples of $t$ are taken in the interval $[10,200]$ years. Computing $P(C(l,t))$ requires a slightly modified version of AMS, where the trajectories not only stop when they reach either the on-state or the collapsed state of the AMOC, but also when they reach a time limit. This version of AMS needs to be run for each of the $20$ values of $t$, from which we can deduce all values of $P(C(l,t))$, as explained in Sect. \ref{sec:ams}. AMS is run $30$ times, using $N=1000$ trajectories, $10$ of which are deleted and replaced at each iteration ($n_c=10$).

\begin{figure}
    \centering
    \includegraphics[width=\textwidth]{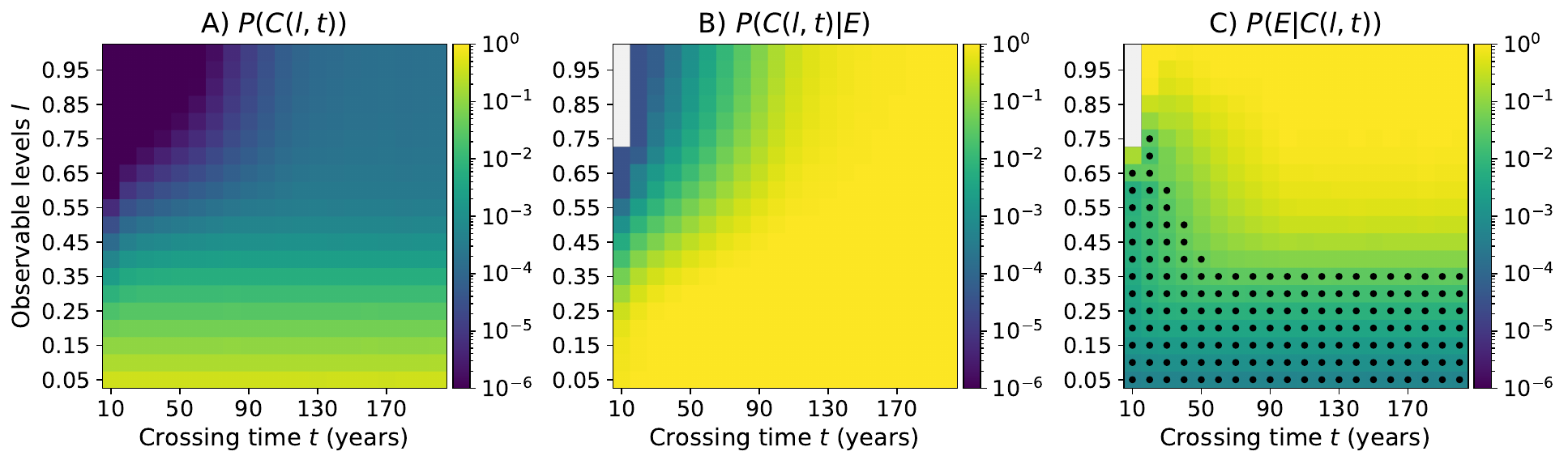}
    \caption{{\it A) Probability of the event $C(l,t)$ for $20$ samples of $l\in[0,1]$ and $20$ samples of $t\in[10,200]$ years.
    B) Probability of the event $C(l,t)$ conditioned on the realization of $E$, for the same samples of $l$ and $t$. This probability is estimated by counting the realizations $C(l,t)$ in $29869$ trajectories obtained from the $30$ runs of AMS performed to compute Fig. \ref{fig:diagram_non_auton}.
    C) Probability of $E$ conditioned on $C(l,t)$ for the same samples of $l$ and $t$. The black dots represent the safe operating space of $E$ conditioned on $C(l,t)$, i.e. the area where $P(E|C(l,t))\leq\alpha$ with $\alpha=0.05$.§}}
    \label{fig:safe_space}
\end{figure}

The quantity $P(C(l,t))$ is represented in Fig. \ref{fig:safe_space}A. As expected, the probability of passing each level $l$ of $\mathcal{O}$ increases with the maximum crossing time $t$. For the lowest levels, closest to the on-state, this time horizon does not impact the probability of crossing but as the system drifts closer to the threshold of collapse, the probability of collapse within the shortest and the longest time horizons differ by several orders of magnitude. 
Fig. \ref{fig:safe_space}B shows the conditional probability $P(C(l,t)\ |\ E)$, which is estimated using all trajectories reaching a collapse among the $3\times10^4$ available after running AMS $30$ times to compute $P(E)$. AMS is stopped as soon as at least $N-n_c+1=991$ trajectories collapse \cite[]{Rolland2022} so we conserved $29869$ trajectories, or $99.56\%$ of the total. The probability to reach the lowest levels is very close to $1$ whatever the value of $t$ but decreases as the system moves closer to the collapse, especially for small $t$, as expected. When the trajectories have no time constraint, their mean first-passage time at the collapse threshold is $123$ years, so the values of $P(C(l,t)\ |\ E)$ are very close to $1$ if not for $t\geq130$ years. In the top-left corner of the diagram, the six grey grid cells correspond to such extreme events (strong decrease of the AMOC strength in less than $10$ years) that they were not found in the studied ensemble and the corresponding probabilities could not be sampled.
Finally, Fig. \ref{fig:safe_space}C shows the posteriori probability of interest: $P(E\ |\ C(l,t))$. The top row of this diagram only contains $1$: the probability of collapsing given that the system collapses within $t$ is $1$. For the rest, each grid cell $(t,l)$ gives the probability that the AMOC collapses before returning to the on-state, if the AMOC strength has been measured and has crossed the level $l$ of the observable within a time $t$ since the last time it left the neighbourhood of the on-state. The lowest value of $P(E\ |\ C(l,t))$ is at the bottom-left grid cell and is $P(E\ |\ C(0.1,10))=4.5\times10^{-4}$, which is more than a factor two with respect to $P(E)=2.0\times10^{-4}$. We can conclude that the knowledge of $C(l,t)$ even for small values of $l$ or $t$ has a significant impact on the assessment of the risk of occurrence of $E$. 

The safe operating space for $E$ conditioned on $C(l,t)$, called $\mathcal{S}$ is defined with respect to a threshold $\alpha$ by:
\begin{equation}
    \mathcal{S}(E,C;\alpha) = \left\{(l,t)\in\mathbb{R}^2\ |\ P(E|C(l,t))\leq\alpha \right\}
\end{equation}
The conditional safe operating space $\mathcal{S}$ is indicated by Fig. \ref{fig:safe_space} by the black dots, where we take $\alpha=0.05$. For $t\leq50$ years, the maximum value of $l$ belonging to $\mathcal{S}$ increases as $t$ decreases, but for $t>50$ years, $\mathcal{S}$ corresponds to all values of $(l,t)$ such that $l\leq0.35$. It means that, for instance, observing the system cross the level $l=0.65$ only $10$ to $20$ years after leaving its statistical equilibrium cannot be interpreted as a strong warning signal for the collapse. It may simply be due to an extreme noise event. For $t>50$ years, the posteriori probability of $E$ does not depend on $t$ anymore, so the largest level of $l$ crossed by the system is the only relevant information to indicate a forthcoming collapse. Any observation of the state of the system below $l=0.35$ cannot indicate a collapse with more than $5\%$ probability. 

\section{Summary and Discussion}
\label{sec:discuss}

We have presented a new framework to compute resilience of the AMOC, for which traditional measures may not be applicable, especially if models become complex. The idea behind our framework is to monitor an observable while the AMOC is losing resilience and transitioning towards a collapsed state in order to characterise its resistance to the forcing that triggers this transition. The ensemble of trajectories needed to conduct this analysis is obtained by running the AMS algorithm, which also allows us to sample information on resilience at no additional computational cost. The background state out of which the system is perturbed and the undesirable region corresponding to a complete loss of resilience can be any region of the phase space but must be defined in accordance with the observable; in that sense our work builds on the generic approach by \cite[]{Schoenmakers2021}. Our contribution, though, expands this idea by proposing a method to compute a resilience measure that directly depends on the chosen observable, which makes it more practical and flexible. 
In particular, we extract two footprints of resilience, related to the ability of the system to return to its background state, and to the resistance it opposes to the imposed forcing. These footprints are then used to build a diagram describing the history of the collapse of the AMOC. This diagram also provides a visual interpretation of resilience, from which we derive a resilience measure ${\cal R}$. The latter is easy to interpret: the larger $\mathcal{R}$ the larger the resilience of the system, the more able it is to resist and counteract the imposed forcing. 
We also measured the resilience of the AMOC for a number of different forcing and noise amplitudes using a conceptual model and were able to quantify how much a change in these values would endanger the AMOC. We showed that our framework is also able to handle non-autonomous systems, which are of increasing importance due to the anthropogenic forcing imposed on the climate system. 
Finally, we extended our framework to define conditional safe operating spaces. Because it is already built on conditional probabilities, our framework can naturally consider the impact of several scenarios given arbitrary conditions. For instance, the influence of greenhouse gas emissions on the resilience of the AMOC could be studied until a certain time-horizon.

Although we focus here on the AMOC, it is important to note that our framework is system-agnostic. AMS can be applied on any system of (theoretically) any dimension, as long as the initial and target domains are well-separated in phase space. The regions considered in this system are also not required to be attractors or have any special properties. The only difficulty may be to find a suitable observable but if the transition between these regions is physically possible, physical arguments may suffice to define an observable capturing some properties of the transition. In particular, we believe that our framework may be useful to bridge the gap between observations and models of any complexity. Moreover, this framework may be useful to compare the resilience of states in different models. Indeed, since this framework is model-independent and only depends on a (normalized) physical observable, the differences in models' dynamics would show in the $(\rho,\Delta\tau)$ diagram. This may be useful to pinpoint where in the transition process differences lie and to better understand these discrpancies, especially in very large models that may be difficult to interpret. 

The notion of critical slowing down \cite[]{Dakos2008,Lenton2012,Scheffer2009} has received much attention in the field of climate, especially to quantify the resilience of the Amazon rainforest \cite[]{Boulton2022} or the AMOC \cite[]{Boers2021,Ditlevsen2023}. It relies on classical statistical quantities such as the increasing variance and auto-correlation that a system should undergo if it approaches a tipping point, and on the restoring rate that can be extracted. However, these measures also rely on a number of assumptions, in particular that the system approaches a bifurcation-induced tipping and that it is subject to Markovian noise. Recent works relax these assumptions \cite[]{Kuehn2022,Morr2024} or provide new results by considering a combination of non-autonomous and noisy perturbations \cite[]{Ritchie2016,Slyman2023}. By comparison, one of the key points of our framework is that it does not rely on any assumption but rather on the efficient sampling of trajectories all undergoing a dramatic change in an observable known to be important for the system. In that way, our measure of resilience is built closer to the underlying dynamics of the system and comes with clear interpretability. This makes it a rather prognostic approach, whereas classical indicators are rather diagnostic.

AMS still comes with a heavy computational cost, due to the large number of trajectories that must be simulated before the algorithm converges, although it is much cheaper and better targeted than blindly computing trajectories using Monte-Carlo simulations. AMS (or algorithms in the same family) is now applied to models of increasing complexity \cite[]{Baars_2021}  including Earth Systems of Intermediate Complexity (EMICs) \cite[]{Cini2024}, which have of the order of $10^5$ degrees of freedom. However, in the latter case, it is  GKTL \cite[]{Giardina2006, Giardina2011}, another algorithm in the same family that was used, whose cost is comparable to AMS but which is less adapted to the sampling of resilience footprints as we describe them here. The TAMS algorithm \cite[]{Lestang_2018,Baars_2021} is a more recent version of AMS but it specifies an end-time for the simulations and does not prevent return to the initial domain. As a result, the transition probabilities are conditioned on this end-time rather on not returning to the initial domain, making TAMS a less suitable algorithm for resilience studies. A limiting aspect of AMS is the choice of score function, because it drives the reliability and the error around the resilience footprints. Unfortunately, there is no general method to design a good score function, although machine-learning methods have recently given promising results \cite[]{Lucente2022,Jacques-Dumas2024,Strahan2023}. 

We  defined the resilience of the system from the characteristic behaviour of an observable as the system undergoes a transition. In other words, the complex dynamics of the model are projected onto a one-dimensional observable, from which it is easier to extract usable footprints of resilience. Of course, information is lost in the process but, the more faithful the observable to the shift of regime, the more reliable its results. The main issue with such single-dimensional observables is that states lying on a given isoline of this observable may describe very different physical situations. As a consequence, estimates of $\rho$ and $\Delta\tau$ along these sole isolines may have a very large variance. A next step would then be to extend this framework to multi-dimensional observables, which would be equivalent as projecting the dynamics of the system in a space of lower dimension. 

Our framework should be connected to actual measurements of the AMOC, to give an estimate of the change in resilience of the actual circulation over the last decades. This could be done for instance by comparing observations or reanalysis data to the predictions of the studied model. In this case, a resilience measure could be defined as a variant of ${\cal R}$ where the integration starts at a certain observation and then follows the simulation from the model. Finally, an interesting application of the conditional safe operating spaces is that of cascading tipping \cite[]{Brummitt2015,Rocha2018,Dekker2018,Klose2021}, where disruption of a leading climate subsystem may trigger the tipping of a following climate subsystem. In the case of the AMOC, its connection with the northern Subpolar Gyre (SPG) is an active area of research \cite[]{Caesar2017,Swingedouw2021,Madan2024}: it is thought that the shutdown of the convection in the SPG may be a precursor to an AMOC collapse. Our framework for safe operating spaces could then compute the probability that an AMOC index declines in the near-future, conditioned on the observation of another index describing the SPG, thus assessing the likelihood of this cascade of events. 

\begin{acknowledgments}
This project has received funding from the European Union's Horizon 2020 research and innovation programme under the Marie Sklodowska-Curie grant agreement no. 956170. H.A. Dijkstra received funding from the European Research Council through the ERC-AdG project TAOC (project 101055096, PI: Dijkstra). C. Kuehn received funding from the ClimTip project; ClimTip has received funding from the European Union’s Horizon Europe research and innovation programme under grant agreement No. 101137601.
\end{acknowledgments}

\section*{Conflict of Interest}
The authors have no conflicts to disclose.

\section*{Competing interests}
The authors have no competing interest.

\section*{Author Contributions}
\textbf{Valerian Jacques-Dumas}: conceptualization (lead), software (lead), formal analysis (lead), writing - original draft (lead), writing - review and editing (equal). \textbf{Henk A. Dijkstra}: conceptualization (supporting), formal analysis (supporting), writing - original draft (supporting), writing - review and editing (equal). \textbf{Christian Kuehn}: conceptualization (supporting), formal analysis (supporting), writing - review and editing (equal). 

\section*{Data Availability Statement}
The Python implementation of the AMS algorithm, the AMOC model, the resilience framework and the code that produces the result plots can be found at the following address: 

https://doi.org/10.5281/zenodo.12608839 \cite[]{code}.

\newpage

\appendix*
\section{AMS algorithm}
\label{app:ams}

First, the initial domain $\mathcal{A}$ and the target domain $\mathcal{B}$ have to be defined. A suitable score function $\Phi$ must be chosen, that obeys certain constraints \cite[]{Baars_2021,Lucente2022} based on these domains. Then, AMS only depends on the number $N$ of simulated ensemble members and the number $n_c$ of trajectories to discard at each iteration. 

\begin{enumerate}
	\item Set $k=0$ and $w_0=1$. 
	\item Simulate $N$ trajectories $\left(\mathbf{X}^{(1)},\dots,\mathbf{X}^{(N)}\right)$ starting in domain $\mathcal{A}$ until they reach domain $\mathcal{B}$ or come back to $\mathcal{A}$. Repeat steps $3-8$ until at least $N-n_c$ (or $N$ if $n_c=1$) trajectories have reached $\mathcal{B}$.
	\item Compute the score function $\Phi(\mathbf{x})$ for every point of each trajectory.
	\item $Q_i$ designates the maximum value of the score function along trajectory $i$. Let $I_k=\{i\in[1,N]\ |\ Q_i\in\{Q^*_{1},...,Q^*_{n_c}\}\}$, where $\{Q^*_j\}_{j\in[1,n_c]}$ are the $n_c$ lowest values of $Q_i$. $I$ may contain more than $n_c$ elements due to the discretization of the trajectories. 
    \item Set $k=k+1$ then $w_k = \left(1-\frac{\#I_k}{N}\right)w_{k-1}$. For every $i\in I_k$, repeat steps $6-8$.
	\item Select a trajectory $\mathbf{X}^{(r)}$ at random, such that $r\notin I_k$. Let $\tau$ the first time so that $\Phi(\mathbf{X}^{(r)}(\tau))\geq Q_i$. 
    \item Set $\mathbf{X}^{(i)}([1,\tau]) = \mathbf{X}^{(r)}([1,\tau])$.
	\item Simulate the rest of $\mathbf{X}^{(i)}$ starting from $\mathbf{X}^{(r)}(\tau)$ until reaching $\mathcal{B}$ or coming back to $\mathcal{A}$.
\end{enumerate}

Let $N_B$ ($N-n_c+1\leq N_B\leq N$) the number of trajectories that reached $\mathcal{B}$ when AMS has terminated and $K$ the number of iterations. The estimated probability of reaching $\mathcal{B}$ starting from $\mathcal{A}$ before returning to $\mathcal{A}$ is given by 
\citep{Rolland2015,Rolland2022}:
\begin{equation}
\label{eq:proba}
    \hat{\alpha} = \frac{N_B}{N}w_K = \frac{N_B}{N}\prod_{k=1}^K w_k = \frac{N_B}{N}\prod_{k=0}^{K-1} \left(1-\frac{\#I_k}{N}\right)
\end{equation}

AMS returns the transition probability from $\mathcal{A}$ to $\mathcal{B}$ after at least $N-n_c+1$ trajectories have reached $\mathcal{B}$. This has to be taken into account when using this algorithm within the resilience framework. Let $l$ an intermediate level of the observable $\mathcal{O}$ such that $\mathcal{O}_\mathcal{A}\leq l \leq \mathcal{O}_\mathcal{B}$. The probability $P(l)$ to transition to $l$ from $\mathcal{A}$ can then only be estimated at the end of the iteration when $N-n_c+1$ trajectories have reached $l$. The mean-first-passage time at this level can then be extracted from the current ensemble of trajectories. 

\newpage

\bibliography{biblio.bib}

\end{document}